
\magnification=1200
\vsize=7.5in
\hsize=5.6in
\tolerance 500
\def\svec#1{\skew{-2}\vec#1}
\def\thru#1{\mathrel{\mathop{#1\!\!\!/}}}
\baselineskip 12pt plus 1pt minus 1pt
\pageno=0
\centerline{\bf TWIST-FOUR DISTRIBUTIONS IN A TRANSVERSELY-POLARIZED NUCLEON}
\smallskip
\centerline{{\bf AND THE DRELL-YAN PROCESS}
\footnote{*}{This work is supported in part by funds provided by the
U. S. Department of Energy (D.O.E.) under contract  \#DE-AC02-76ER03069.}}
\vskip 24pt
\centerline{{Pervez Hoodbhoy}
\footnote{\dag}{Supported by NSF Grant INT-9122027.}}
\vskip 12pt
\centerline{\it Center for Theoretical Physics}
\centerline{\it Laboratory for Nuclear Science}
\centerline{\it and Department of Physics}
\centerline{\it Massachusetts Institute of Technology}
\centerline{\it Cambridge, Massachusetts\ \ 02139\ \ \ U.S.A.}
\centerline{\it and}
\centerline{\it Department of Physics}
\centerline{\it Quaid-e-Azam University}
\centerline{\it Islamabad, Pakistan}
\vskip 24pt
\centerline{Xiangdong Ji}
\vskip 12pt
\centerline{\it Center for Theoretical Physics}
\centerline{\it Laboratory for Nuclear Science}
\centerline{\it and Department of Physics}
\centerline{\it Massachusetts Institute of Technology}
\centerline{\it Cambridge, Massachusetts\ \ 02139\ \ \ U.S.A.}
\vfill
\baselineskip 24pt plus 2pt minus 2pt
\centerline{\bf ABSTRACT}
\medskip
The twist-four quark and gluon distributions in a transversely polarized
nucleon are identified and relations among them are discussed by using QCD
equations of motion.  Their contribution, at the order of
${\cal O}\left(1/Q^2\right)$,
to the Drell-Yan cross section in transversely
polarized nucleon-nucleon collisions is calculated.
\vfill
\centerline{Submitted to: {\it Phys. Rev. D}}
\vfill
\line{CTP\#2203 \hfil April 1993}
\eject

\noindent{\bf I.\quad INTRODUCTION}
\medskip
\nobreak
There has been considerable interest recently
in the polarization-dependent quark distributions
of a transversely polarized nucleon.$^{1-4}$
These distributions, which are
on the same footing as those of an unpolarized or
longitudinally polarized nucleon, are just as important
for characterizing the nucleon's high-energy structure.
Being universal quantities --- meaning that they describe the
nucleon rather than a particular process --- these
distributions can be measured in a variety
of experiments involving large momentum transfers.
Although at present no parton distribution can be calculated
ab-initio from QCD, nonetheless this may
eventually become possible.

A very interesting distribution, $h_1(x)$, called
the transversity distribution in ref. 2,
was first identified by Ralston and Soper.  That this
involves a helicity flip of the participating quark
was recognized by Artru and Mekhfi.$^{1}$  The $h_1(x)$
is a twist-two quantity, meaning that it enters into the cross
section of a hard
process, such as Drell-Yan lepton pair production, unsuppressed
by inverse powers of the hard momentum $Q\ $.  It has been
proposed that $h_1(x)$ be
measured using transversely polarized beams at RHIC.$^{4}$  The ${\cal
O}\left(1\over Q^2\right)$, or twist-four, corrections to
this process involve parton distributions with four partons.
The purpose of this paper is to identify and classify these
distributions and to calculate their contribution to the
Drell-Yan process.

Higher twist corrections, as is well-known, are notoriously complicated.
Nonetheless, if one is to proceed beyond the naive parton model, it is
necessary to identify all parton distributions which enter at higher
twist, to study their symmetry properties, and to calculate their
contributions to different hard processes.  In contrast to simple
distributions involving only two quark or gluon fields
which occur at leading twist,
higher twist distributions involve the matrix elements
of the multi quark and gluon field operators
at equal light-cone times.
A complete catalogue of unpolarized and longitudinally
polarized distributions up to
twist-four, and transverse distributions up to twist-three,
can be found in ref.~3.

The results of this paper, which deals exclusively
with the twist-four distributions of a transversely
polarized nucleon, are summarized below:

Parton distributions are introduced through
parton-hadron vertices. As is shown in Fig. 1,
the twist-four distributions in the light-cone gauge
are contained in the vertices up to four partons.
Here the four gluon vertex is absent because it
cannot mix with the {\it chiral-odd} vertices
shown. The two-quark vertex, shown in fig.~1a, provides
one twist-four distribution $h_3(x)$.
There are four distributions, named $d_i(x,y)$ with
$i=1,2,3,4\ $, associated with the two-quark-one-gluon vertex
in fig.~1b.  The two-quark-two-gluon vertex, fig.~1c, gives
rise to three distributions named herein
as $D_i(x,y,z)$ with $i=1,2,3\ $.  Finally, the four-quark
vertex (fig.~1d) has two associated distributions,
$W_i(x,y,z)$ with $i=1,2\ $.  The tensor
structure which gives rise to each distribution is given in the text, together
with restrictions obtained from the requirement of PCT invariance.  The QCD
equations of motion on the light-cone impose further restrictions by
specifying relations between distributions involving more light-cone momentum
fractions and distributions involving fewer.  Including the distributions
identified in ref.~3 --- i.e.~$h_1(x)\ ,\ g_T(x)\ ,\ G_1(x,y)\ ,\ {\rm and}\
G_2(x,y)$ --- a complete set is now available to deal with any hard process
involving transversely polarized nucleons up
to and including twist-four.  As one application, we have calculated the
twist-four part of the Drell-Yan cross section for transversely polarized
nucleons.  This could provide a framework to analyze corrections if an
experiment to measure $h_1(x)$ is actually performed$^{4}$.

\bigskip
\goodbreak
\noindent{\bf II. TWIST-FOUR DISTRIBUTIONS IN A
TRANSVERSELY POLARIZED NUCLEON}
\medskip
\nobreak

In this section, we consider the polarization-dependent
twist-four distributions in a transversely polarized nucleon.
To establish notation, we consider a nucleon moving in the $z$ direction with
its spin vector in the $x-y$ plane with momentum $P^\mu=p^\mu+{1\over
2}M^2n^\mu$ and $S^\mu_\perp=(0,{\svec S}_\perp,0)\ $, where ${\svec
S}_\perp\cdot{\svec S}_\perp=1$ and $p^\mu$ and $n^\mu$ are null vectors
satisfying $p^2=n^2=0$ and $p\cdot n=1\ $.  A distribution of partons in such
a nucleon is defined by the matrix element of quark and gluon
field operators at equal light-cone
times.  We choose the light-cone gauge, $A\cdot n = 0$,
Simple dimensional reasoning enables determination of the number of
inverse powers of the hard momentum which accompany a given distribution, and
thus its twist.  Application of PCT rules out a large number of otherwise
possible structures.  For further details the reader is referred to refs.~2
and 3.

At the level of twist-four, the two quark vertex in fig.~1a
contains only one transverse distribution,
$$h_3(x)={1\over \Lambda^2}\int{d\lambda\over 2\pi}e^{i\lambda x}\left\langle
PS_\perp\left\vert{\bar\psi}(0)\gamma_5\thru S_\perp\thru
p\psi(\lambda n)\right\vert PS_\perp\right\rangle\ ,\eqno(1)$$
where $\Lambda$ is a soft scale. It can also be
projected out from the quark density matrix,
$$\eqalign{
M(x) &=\int{d\lambda\over 2\pi}e^{i\lambda x}\left\langle
PS_\perp\left\vert{\bar\psi}(0)\psi(\lambda n)\right\vert
PS_\perp\right\rangle \cr
 &={1\over 4}\Lambda^2{\thru n}\gamma_5{\thru S}_\perp h_3(x)+...\
,\cr}\eqno(2)$$
where dots represent other distributions which are
our concern here.

The two-quark-one-gluon vertex, fig.~1b, is more complex and involves two
light-cone
fractions since the quark and gluon can be removed from the hadron from
different light-cone ``positions'' $x^-=\lambda n^-$ or $y^-=\mu n^-\ $.
There are four PCT allowed transverse distributions:
$$\eqalignno{
d_1(x,y) &=\int[d\lambda\, d\mu]\left\langle
PS_\perp\left\vert{\bar\psi}(0)iD_\perp(\mu n)\cdot
S_\perp\gamma_5\psi(\lambda n)\right\vert PS_\perp\right\rangle\ , &  \cr
d_2(x,y) &=\int[d\lambda\, d\mu]\left\langle
PS_\perp\left\vert{\bar\psi}(0)iD_\perp(\mu n)\cdot S_\perp\gamma_5{1\over
2}(\thru p\thru n-\thru n\thru p)\psi(\lambda n)\right\vert PS_\perp
\right\rangle\ , & \cr
d_3(x,y) &=\int[d\lambda\, d\mu]\left\langle
PS_\perp\left\vert{\bar\psi}(0)iD_\perp(\mu n)\cdot
iT_\perp\psi(\lambda n)\right\vert PS_\perp\right\rangle\ , &  \cr
d_4(x,y) &=\int[d\lambda\, d\mu]\left\langle
PS_\perp\left\vert{\bar\psi}(0)iD_T(\mu n)\cdot iT_\perp{1\over 2}(\thru
p\thru n-\thru n\thru p)\psi(\lambda n)\right\vert PS_\perp\right\rangle\ , &
({\rm 3}) \cr}$$
where $T_\perp^\alpha=\epsilon^{\alpha\beta\gamma\delta}p_\beta n_\gamma
S_{\perp\delta}$ is a vector orthogonal to $S^\alpha_\perp$ with $T_\perp\cdot
T_\perp=-1$ and,
$$[d\lambda\, d\mu]={1\over 2\Lambda^2}{d\lambda\over 2\pi}{d\mu\over
2\pi}e^{i\lambda
x}e^{i\mu(y-x)}\ . \eqno(4)$$
The $d_i(x,y)$ may be readily projected out of the quark-gluon
density matrix,
$$\eqalignno{
M^\alpha(x,y) =&2\Lambda^2\int[d\lambda\, d\mu]\left\langle
PS_\perp\left\vert{\bar\psi}(0)iD^\alpha_\perp(\mu
n)\psi(\lambda n)\right\vert PS_\perp\right\rangle & \cr
 =&-{1\over 2}\Lambda^2\Big[S^\alpha_\perp\gamma_5d_1(x,y)+
S^\alpha_\perp{1\over 2}(\thru p\thru n-\thru n\thru p)\gamma_5d_2(x,y) &
\cr
 &- iT^\alpha_\perp d_3(x,y)- iT^\alpha_\perp{1\over
2}(\thru p\thru n-\thru n\thru p)d_4(x,y)\Big] +... & (5) \cr}$$
It can be easily shown that the $d_i$'s are real and obey the symmetry
relations:
$$\eqalign{
d_1(x,y) &=-d_1(y,x)\ ,\ d_2(x,y)=d_2(y,x) \cr
d_3(x,y) &=-d_3(y,x)\ ,\ d_4(x,y)=d_4(y,x)\ . \cr} \eqno(6) $$
Note that the index $\alpha$ in eq. (5) refers to transverse polarizations
$(\alpha=1,2)$ only because $A^+=0$ and $A^-$ is a dependent quantity whose
presence would make a distribution belong to one higher twist.

The two-quark and two-gluon vertex, fig.~1c, also with
transverse gluons only, has three
light-cone fractions.  There are three allowed distributions:
$$\eqalignno{
D_1(x,y,z) = &\int[d\lambda\, d\mu\, d\nu]\left\langle
PS_\perp\left\vert{\bar\psi}(0)iD_\perp(\nu n)\cdot iD_\perp(\mu n)\thru
n\gamma_5\thru S_\perp\psi(\lambda n)\right\vert PS_\perp\right\rangle\ , &
 \cr
D_2(x,y,z) = &\int[d\lambda\, d\mu\, d\nu]\Big\langle
PS_\perp\big\vert{\bar\psi}(0)[iD_\perp(\nu n)\cdot S_\perp i\thru D_\perp(
\mu n)+i\thru D_\perp(\nu n)iD_\perp(\mu n)\cdot
S_\perp & \cr
 &-iD_\perp(\nu n)\cdot iD_\perp(\mu n)\thru S_\perp]\thru
n\gamma_5\psi(\lambda n)\big\vert PS_\perp\Big\rangle\ , &  \cr
D_3(x,y,z) = &\int[d\lambda\, d\mu\, d\nu]\left\langle
PS_\perp\left\vert{\bar\psi}(0)i\epsilon^{\alpha\beta\gamma\delta}p_\gamma
n_\delta iD_{\perp\alpha}(\nu n)iD_{\perp\beta}(\mu n)\thru n\thru
S_\perp\psi(\lambda n)\right\vert PS_\perp\right\rangle & ({\rm 7}) \cr}$$
where,
$$[d\lambda\, d\mu\, d\nu]={1\over 2\Lambda^2}{d\lambda\over 2\pi}{d\nu\over
2\pi}{d\mu\over 2\pi}e^{i\lambda x}e^{i\mu(y-x)}e^{i\nu(z-y)}\ . \eqno(8)$$
The above distributions can be projected out of the $qGGq$ density matrix,
$$\eqalignno{
M^{\alpha\beta}(x,y,z)= &2\Lambda^2\int[d\lambda\, d\mu\, d\nu]\left\langle
PS_\perp\left\vert{\bar\psi}(0)iD^\alpha(\nu n)iD^\beta(\mu n)\psi
(\lambda n)\right\vert PS_\perp\right\rangle & \cr
= &\ {1\over 4}\Lambda^2\Big[g^{\alpha\beta}_\perp\gamma_5\thru S_\perp\thru
pD_1(x,y,z) &
\cr
 &+{1\over 2}\left(S^\alpha_\perp\gamma^\beta+S^\beta_\perp\gamma^\alpha
-g^{\alpha\beta}_\perp\thru S_\perp\right)\thru p\gamma_5D_2(x,y,z) & \cr
 &+i\epsilon^{\alpha\beta\gamma\delta}p_\gamma n_\delta\thru
S_\perp\thru pD_3(x,y,z)\Big] +... & (9) \cr}$$
Working from the definitions 7a-7c, certain symmetry relations can be
established,
$$\eqalignno{
D_1(x,y,z) &=D_1(z,y,x), &  \cr
D_3(x,y,z) &=-D_3(z,y,x)\ . & (\hbox{10}) \cr} $$

Finally, the four quark matrix element in fig.~1d can be similarly analyzed.
The twist-four transverse spin distributions are:
$$\eqalignno{
W_1(x,y,z) &={g^2\over 2}\int[d\lambda\, d\mu\, d\nu]\left\langle
PS_\perp\left\vert{\bar\psi}(0)\thru n\psi(\nu n){\bar\psi}(\mu n)\thru
n\gamma_5\thru S_\perp\psi(\lambda n)\right\vert PS_\perp\right\rangle\ , &
 \cr
W_2(x,y,z) &={g^2\over 2}\int[d\lambda\, d\mu\, d\nu]\left\langle
PS_\perp\left\vert{\bar\psi}(0)\thru
n\gamma_5\psi(\nu n){\bar\psi}(\mu n)\thru n\thru
S_\perp\psi(\lambda n)\right\vert PS_\perp\right\rangle\ . & (\hbox{11})
\cr}$$
The four-quark density matrix, which is a matrix in two sets of Dirac indices,
from which these can be projected is,
$$\eqalignno{
M(x,y,z)= &2g^2\Lambda^2\int[d\lambda\, d\mu\, d\nu]\left\langle
PS_\perp\left\vert{\bar\psi}(0)\psi(\nu n){\bar\psi}(\mu n)\psi
(\lambda n)\right\vert PS_\perp\right\rangle & \cr
= &{1\over 4}\Lambda^2\Big[(\thru p)(\gamma_5\thru S_\perp\thru p)W_1(x,y,z)
+ (\gamma_5\thru p)(\thru p\thru S_\perp)W_2(x,y,z)\Big] + ... & (12) \cr}$$

At this point one needs to take stock of the situation.  So far we have, using
PCT, Lorentz invariance, and dimensional counting, identified all tensor
structures needed to describe a transversely polarized nucleon at the
twist-four level.  However, we have further constraints available to us in the
form of QCD equations of motion on the light-cone,
$$i{d\over d\lambda}\psi_-(\lambda n)=-{1\over 2}\thru ni\thru
D_\perp\psi_+(\lambda n) $$
$$\psi_\pm={1\over 2}\gamma^\mp\gamma^\pm\psi\ . \eqno(\hbox{13})$$
This can be used to establish relations between distributions with three
light-cone fractions with those having two and one, etc.  A particularly useful
set of relations, which shall be needed for establishing the electromagnetic
gauge invariance of the Drell-Yan cross is,
$$\eqalignno{
\int dz\, D_2(x,y,z) &=-y(d_1(x,y)+d_2(x,y)+d_3(x,y)+d_4(x,y))\ , &\cr
\int dz\, D_2(z,x,y) &=x(d_1(x,y)+d_3(x,y)-d_2(x,y)-d_4(x,y))\ , &\cr
\int dy\, dz\, D_2(y,x,z) &=x^2h_3(x)\ . & (\hbox{14}) \cr}$$

All distributions discussed so far are implicitly dependent on
the renormalization scale. Their evolution through radiative
processes, however, is a
complicated issue and beyond the scope of this paper.

\bigskip
\goodbreak
\noindent{\bf III. THE TRANSVERSELY POLARIZED DRELL-YAN PROCESS}
\medskip
\nobreak
We now consider application to the Drell-Yan production of lepton pairs in
transversely polarized nucleon-nucleon
collisions.  Calculations beyond
leading twist require considerable formal development,
as in the work of Ellis, Furmanski,
and Petronzio, and more recently by Qiu and Sterman,
and Jaffe and Ji.$^{2,5,6}$  A
collinear expansion of the parton momenta is carried out,
different Feynman diagrams are combined together
to arrive at colour gauge invariance, etc.
However, a simpler set of rules emerges from the formalism, which is
summarized in ref.~3.  These rules may be readily applied to
the Drell-Yan
process by first drawing the set of all diagrams which give rise to a lepton
pair in the final state, and which do not contain more than four partons
belonging to a single nucleon.  The set of diagrams contributing up to
twist-four are given in figs.~2-5.
All diagrams of a given topology
must be included, although only one has actually been shown in the figures.

The hadron tensor that
appears in the inclusive Drell-Yan cross section can be
written as,
$$W^{\mu\nu}=\int dx\, dy(2\pi)^4\delta^4(Q-xp_A-yp_B)\Omega^{\mu\nu}(x,y)\ ,
\eqno(\hbox{15})$$
where $x(>0)$ and $y(>0)$ are the momentum fractions carried by quarks
or antiquarks from the nucleon $A$ and $B$, respectively.
For our purpose, we are interested in only the polarization-dependent
part of the tensor. In the following discussion,
we assume that a quark from $A$
annihilates with an antiquark from $B$.
for the opposite case,
an antiquark from $A$ annihilating a quark from $B$,
is obtained by substituting
$x\rightarrow -x$ and $-y\rightarrow x$.

To give an example of the application of the rules in ref. 3,
we calculate the contribution from fig. 2a,
$$\Omega^{\mu\nu}(x,y)|_{2a}=(-1){\rm Tr}\gamma^\nu M_A(x)\gamma^\mu M_B(-y)\ .
\eqno(\hbox{16})$$
The factor $(-1)$ arises from the anti-commuting nature of the fermion fields,
and an implicit trace over colour is understood.
$M_A(x)$ and $M_B(-y)$ are the transverse-polarization-dependent
part of the quark density matrix,
$$    M(x) = {1\over 2}\gamma_5\thru S_\perp \thru p h_1(x)
             + {1\over 2} \Lambda \gamma_5
            \thru S_\perp g_T(x) +
             {1\over 4}\Lambda^2 \thru n \gamma_5\thru S_\perp h_3(x).
\eqno(17) $$
Working out the fermion and colour traces, we find
$$\eqalignno{\Omega^{\mu\nu}(x,y)|_{2a}
 = & {1\over 4}C_F\Lambda^2(S_{\perp A}\cdot S_{\perp B})
\Big[P^\mu_BP^\nu_Bh_1^B(-y)h_3^A(x) \cr & \  +
P^\mu_AP^\nu_Ah_3^B(-y)h_1^A(x)\Big]
{1 \over P_A\cdot P_B} & \cr
 - &{1\over 4}C_F\Lambda^2 \Big[{\cal S}^{\mu\nu}
   - S_{\perp A}\cdot S_{\perp B}{P_A^\mu P_B^\nu
   + P_A^\nu P_B^\mu \over P_A\cdot P_B}\Big] g^B_T(-y)g^A_T(x), & (18)
 \cr}$$
where
$${\cal S}^{\mu\nu}=S_{\perp A}^\mu S_{\perp B}^\nu
         + S_{\perp A}^\nu S_{\perp B}^\mu
         - g_T^{\mu\nu}S_{\perp A}\cdot S_{\perp B},
    \eqno(\hbox{19})$$
The first term contains only the chiral-odd
distributions and the second the chiral-even
distributions. Quark helicity conservation for massless
quarks dictates that the Drell-Yan cross section consists
only of the combinations of chiral odd -- chiral
odd distributions or chiral even -- chiral even
distributions. $C_F=(N^2-1)/2N=4/3$
is a colour factor for three colours.

Fig.~2b contains one gluon coming from
the nucleon $A$, and the corresponding expression
for $\Omega^{\mu\nu}$ is,
$$\Omega^{\mu\nu}(x,y)|_{2b}=(-1)\int dz\, {\rm Tr}\left[\gamma^\nu
M^\alpha_A(z,x)\gamma^\mu M_B(-y)i\gamma_\alpha {i\over (z-x)\thru p_A-y\thru
p_B}\right]\ . \eqno(\hbox{20})$$
where the twist-four part of the matrix $M_A(z,x)$ is given
in eq.~(5), and the twist-three part in eq. (43) in ref. 3.
After a lengthy calculation, we find the contributions
from all four diagrams having the fig. 2b topology,
$$\eqalignno{ \Omega^{\mu\nu}(x,y)|_{2b} =
  &-{1\over 4}C_F\Lambda^2
(S_{\perp A}\cdot S_{\perp B})
{(P_A^\mu P_B^\nu+P_A^\nu P_B^\mu)\over P_A\cdot P_B} \cr & \ \ \
\Big[{h_1^B(-y)\over y}xh_3^A(x) + {h_1^A(x)\over x}yh_3^B(-y)\Big] & \cr
 &+{1\over 4}C_F\Lambda^2(S_{\perp A}\cdot S_{\perp B}) g^{\mu\nu}_T
\Big[{h_1^B(-y)\over
y}\int dw\,{dz\over z}D_2^A(x,z,w) \cr & \ \ \ + {h_1^A(x)\over
x}\int dw\,{dz\over z}D_2^B(-y,-z,-w) \Big] & \cr
 &-{1\over 2}C_F\Lambda^2 (S_{\perp A}\cdot S_{\perp B})
\Big[{x^2P_A^\mu P_A^\nu
 + y^2P_B^\mu P_B^\nu \over xyP_A\cdot
P_B}\Big] g_T^B(-y)g_T^A(x)  & \cr
 &+{1\over 2}C_F\Lambda^2 (S_{\perp A}\cdot S_{\perp B})
g_T^{\mu\nu}\Big[g_T^B(-y)\int dz\,{G_1^A(x,z)-G_2^A(x,z)\over x-z}
\cr &  \ \ \
 - g_T^A(x)\int dz\,{G_1^B(-y,-z)-G_2^B(-y,-z)\over y-z}\Big] & (21) \cr} $$
The first two terms in the above formula are chiral odd,
and the second two are chiral even.  The transverse metric
tensor has only $g_T^{11}=g_T^{22}=-1$ as non-zero components.

For the diagrams with two gluons from the same
nucleon shown in figs. 3a and 3b,
there are only chiral-odd contributions.
Using the density matrix in eq. (9), we find,
$$\eqalign {\Omega^{\mu\nu}(x,y)|_{3a}= & {1\over 4}C_F\Lambda^2
(S_{\perp A}\cdot S_{\perp B}) \Big[x^4P_A^\mu P_A^\nu h_1^B(-y)h_3^A(x) \cr &
    + y^4P_B^\mu P_B^\nu h_3^B(-y)h_1^A(x)\Big]{1\over x^2y^2P_A\cdot P_B}, }
\eqno(22)$$
$$\eqalign{ \Omega^{\mu\nu}(x,y)|_{3b} = &
    -{1\over 4}C_F \Lambda^2{\cal S}^{\mu\nu}  \Big[
  {h_1^B(-y)\over y}\int
  dx_1\, dx_2\, {D_1^A(x,x_2,x_1)+D_3^A(x,x_2,x_1)\over x-x_1} \cr & +
 {h_1^A(x)\over x}\int
  dx_1\, dx_2\, {D_1^B(-y,-x_2,-x_1)+D_3^B(-y,-x_2,-x_1)\over y-x_1}\Big]\ , }
\eqno(\hbox{23})$$
In eq. (22), we have made use of eq. (14c). Clearly,
the chiral-odd longitudinal part from eq. (22),
combined with these from eqs. (18) and (21)
is electromagnetically gauge invariant.

There are four diagrams in which one gluon comes from $A$
and the other comes from $B$. These diagrams
contribute only to chiral-even part of the cross section.
Our calculation shows,
$$\eqalignno{
 \Omega^{\mu\nu}(x,y)|_{4a} = &{1\over 4}C_F\Lambda^2
(S_{\perp A}\cdot S_{\perp B})
{P_A^\mu P_B^\nu+P_A^\nu P_B^\mu\over P_A\cdot P_B}
g_T^B(-y)g_T^A(x) & (24) \cr
\Omega^{\mu\nu}(x,y)|_{4b} = &C_F\Lambda^2(S_{\perp A}\cdot S_{\perp B})
g_T^{\mu\nu}\int{dz\,
dw\over z(w-y)}\left[G_1^A(x,w)-G_2^A(x,w)\right] & \cr
 &\ \ \ \ \times \left[G_1^B(-z,-y)
-G_2^B(-z,-y)\right]  + (x\leftrightarrow -y, A\leftrightarrow B)& \cr
 &-{1\over 2}C_A\Lambda^2(S_{\perp A}\cdot S_{\perp B})
g_T^{\mu\nu}\int{dz\, dw\over
z(w-x)}\left[{\tilde G}_1^A(x,w)-{\tilde G}_2^A(x,w)\right] & \cr
 &\ \ \ \ \times \left[{\tilde
G}_1^B(-z,-y)-{\tilde G}_2^B(-z,-y)\right]
  + (x\leftrightarrow -y, A \leftrightarrow B)& (25) \cr}$$
where $C_A = 3$. The second term
in eq. (25) requires some explanation:  ${\tilde G}_i(x,w)$
is defined exactly as $G_i(x,w)$ as in eq.~38 of ref. 3,
except that the covariant derivative
has only the gluon potential. This is because
the partial derivative coming from transverse
momentum expansion of fig. 2a does not contribute
to terms multiplied by $C_A$. Since the final
result has to be gauge invariant, we replace
the $A^\perp$ by $in^-F^{+\perp}/(w-x)$,
and thus, ${\tilde G}_1(x,w)$ is manifestly
gauge invariant. And,
$$\eqalignno{
 \Omega^{\mu\nu}(x,y)|_{4c} =
    &C_F\Lambda^2{\cal S}^{\mu\nu}\int{dz\over z}\,{dw\over w}
        \left[G_1^A(x,w) + G_2^A(x,w)\right]\cr & \ \ \
       \times \left[G_1^B(-y,-z)+G_2^B(-y,-z)\right] & \cr
 &-{1\over 2}C_A\Lambda^2{\cal S}^{\mu\nu}\int {dz\over z}\,{dw\over
w}\left[\tilde G_1^A(x,w)+\tilde G_2^A(x,w)\right] & \cr
 &\  \ \ \times \left[\tilde
G_1^B(-y,-z)+\tilde G_2^B(-y,-z)
\right] & (26) \cr
 \Omega^{\mu\nu}(x,y)|_{4d} = &-C_A\Lambda^2 (S_{\perp A}\cdot S_{\perp
B}) g_T^{\mu\nu}\int{dz\over z-y}\,{dw\over
w-x}\big[\tilde G_1^A(x,w)\tilde G_1^B(-y,-z) & \cr
 &-\tilde G_2^A(x,w)\tilde
G_2^B(-y,-z)\big] & (27)
\cr}$$
Again, it is simple to see that
the chiral-even longitudinal part in eq. (24),
combining with these from eqs. (18) and (21)
is electromagnetically gauge invariant.

Finally, we consider the four-quark diagram
shown in fig.~5.  Evaluation
requires a double trace as there are two quark loops.  This is
straightforwardly done, and using eq.~(12), we get:
$$\eqalignno{
   \Omega^{\mu\nu}(x,y)|_5 =  &-{1\over 4}C_F
       \Lambda^2{\cal S}^{\mu\nu}{h_1^B(-y)\over
2y}\int{dz\over z-x}\,{dw\over w-x}\sum_{i=1}^2\Big[W_i^A(x,z,w) & \cr
 &\ \ \  +W_i^A(x-z,-z,w-z)-W_i^A(x,z,z-w)-W_i^A(x-z,-z,-w)\Big] & \cr
\ \ \ &-{1\over 4}C_F\Lambda^2{\cal S}^{\mu\nu}{h_1^A(x)\over
2x}\int{dz\over z-y}\,{dw\over w-y}\sum_{i=1}^2\Big[W_i^A(-y,-z,-z+w) & \cr
 &\ \ \  +W_i^A(x-z,-z,w-z)-W_i^A(-y,-z,-w) \cr & \ \ \
    -W_i^A(-y+z,z,-w+z)\Big] & (28) \cr}$$
The different terms in eq. (28) comes from eight different
diagrams having the same topology as fig. 5.

It may be readily verified that our final result for $W^{\mu\nu}\ $, which is
given by the sum of all contributions from diagrams 2-5, is
electromagnetically gauge invariant:  $q^\mu W_{\mu\nu}=W_{\mu\nu}q^\nu=0\ $.
This is an important partial check on the correctness of the calculation.

\bigskip
\goodbreak
\noindent{\bf IV. SUMMARY}
\medskip
\nobreak

In this paper, we introduce the
polarization-dependent twist-four distributions
in a transversely polarized nucleon.
The most general ones contain three light-cone
momentum fractions and are $D_i$ and $W_i$
defined in eqs. (7) and (11). There distributions,
together with the twist-two and twist-three
distributions defined for the same nucleon
state, form a complete set for
describing any hard scattering processes
involving traverse polarization.

As an example, we have calculated the twist-four
correction to the Drell-Yan process. The final
result is gauge invariant in both
electromagnetic and colour interactions.
For lack of information on these distributions,
we cannot access numerical significance
of the correction other than simple
dimensional analysis. However, our formulas can be
coupled with model calculations of the
distributions to give a detailed
prediction.

\vfill
\eject

\centerline{\bf REFERENCES}
\bigskip
\item{1)} J.~Ralston and D.E.~Soper, {\it Nucl. Phys.\/} {\bf B152}, l09,
(1979).
\medskip
\item{} X.~Artu and M.~Mekhfi, {\it Z.~Phys.\/} {\bf C45}, 669, (1990).
\medskip
\item{2)} R.L.~Jaffe and X.~Ji, {\it Phys. Rev. Letts.\/} {\bf 67}, 552,
(1991).
\medskip
\item{} R.L.~Jaffe and X.~Ji, {\it Nucl. Phys.\/} {\bf B375}, 527,
(1992).
\medskip
\item{3)} X.~Ji, MIT-CTP preprint \#2141, (1992).
\medskip
\item{4)} G.~Bunce et al, {\it Part. World\/} {\bf 3}, 1, (1992);
\medskip
\item{} X. Ji, {\it Phys. Lett.} {\bf  284}, 137, (1992).
\medskip
\item{5)} R.K.~Ellis, W.~Furmanski, and R.~Petronzio, {\it Nucl. Phys.\/}
{\bf B212}, 29, (1983).
\medskip
\item{6)} J.~Qiu and G.~Sterman, {\it Nucl. Phys.\/} {\bf B353}, 105,
(1991).
\medskip
\vfill
\eject
\centerline{\bf FIGURE CAPTIONS}
\centerline{For a hard copy of the figures, please send email to
ereidell@marie.mit.edu}
\medskip
\item{Fig.~1:} Parton density matrices which can
contribute to a hard process at the twist-four level.
\medskip
\item{Fig.~2:} A quark from hadron A annihilates with an
antiquark from hadron B leading to a large mass virtual
photon. Other diagrams of this type are obtained by allowing
the gluons in (b) to interact with the quark line
on the right, and interchanging A and B.
\medskip
\item{Fig.~3:} Representative diagrams which give chiral
odd contributions to the twist-four Drell-Yan cross section.
\medskip
\item{Fig.~4:} Diagrams involving one gluon from each
nucleon which give chiral even contribution to the
Drell-Yan cross section.
\medskip
\item{Fig.~5:} Four quark diagram contributing to the
Drell-Yan twist-four cross section.

\par
\vfill
\end